\documentclass{PoS}
\pdfoutput=1
\title{Search for R-Parity Violating Supersymmetry at the CMS Experiment}

\ShortTitle{Search for R-Parity violating Supersymmetry}

\author{\speaker{Altan Cakir}\\
       {\rm On behalf of the CMS Collaboration}\\
       Deutsches Elektronen-Synchrotron (DESY)\\
       E-mail: \email{cakir@cern.ch}}

\abstract{The latest results from CMS on R-Parity violating Supersymmetry based on the 19.5/fb full dataset from the 8 TeV LHC run of 2012 are reviewed. 
The results are interpreted in the context of simplified models with multilepton and b-quark jets signatures that have low missing transverse energy arising from light top-squark pair with R-parity-violating decays of the lightest supersymmetric particle. In addition to simplified model, a new approach for phenomenological MSSM interpretation is shown which demonstrates that the obtained results from multilepton final states are valid for a wide range of supersymmetry models.}

\FullConference{The European Physical Society Conference on High Energy Physics \\
		 18-24 July, 2013\\
		 Stockholm, Sweden}

\begin{document}

\section{Introduction}

Searches for supersymmetry (SUSY) have taken an unexpected turn with the Higgs discovery at 125 GeV~\cite{Chatrchyan:2012ufa}. The contributions of SUSY particle loops to the Higgs mass is at most $(m_{h}^{tree})^{2}$ $\le$ $m_{Z}^{2}$, implying top/supersymmetric-top (stop) loops provide the necessary contribution to stabilize the electroweak scale~\cite{Papucci:2011wy}.  Experiments therefore show us that, if SUSY exists, it is either tuned or extended, or it does not fullfill the standard approaches and that more complicated models, with possibility to additional contributions to the model, have to be taken into account. It is well known that in most of the SUSY models, where R-parity is conserved, superpartners can only be produced in pairs and the lightest supersymmetric particle (LSP) is stable, and serves as a a dark matter candidate. In the last decade R-parity violation (RPV) scenarios have been considered as unlikely models for a supersymmetric extension of the Standard Model (SM)~\cite{Barbier:2004ez}. 

R-parity is a discrete symmetry, which can be defined as $R_P$ = $(-1)^{(3B+L+2s)}$. Here $B$ denotes the baryon number, $L$ the lepton number and $s$ the spin of a particle. SUSY models with RPV interactions necessarily violate either B or L but can avoid proton decay limits. The most general RPV superpotential terms can be written as:

\begin{equation}
\label{eq:gen}
W_{RPV} = \lambda_{ijk}L_i L_j \bar{E}_k + \lambda^{'}_{ijk}L_i Q_j \bar{D}_k + \lambda^{''}_{ijk}\bar{U}_i \bar{D}_j \bar{D}_k
\end{equation}
where $i, j$ and $k$ are generation indices; $L$ and $Q$ are the $SU(2)_{L}$ doublet superfields of the lepton and quark; and the $\bar{E}$, $\bar{D}$, and $\bar{U}$ are the $SU(2)_{L}$ singlet superfields of the charge lepton, down like quark, and up-like quark. The third term violates baryon number conservation, while the first and second terms violate lepton number conservation. In the following sections several searches for SUSY based on the leptonic RPV in events with multilepton final states are discussed. All analyses are performed using the full dataset collected with the Compact Muon Solenoid (CMS)~\cite{Ball:2007zza} in proton-proton collisions at a center-of-mass energy of 8 TeV, corresponding integrated luminosity of $19.5$/fb.

\section{Search for top squarks in R-parity-violating supersymmetry using three or more leptons and b-tagged jets}

In this analysis, the result of a search for pair production of top squarks with RPV decays of the lightest sparticle using multilepton events with one or more b-quark tagged jets is presented~\cite{Chatrchyan:2013xsw}. Events with three or more leptons (including tau leptons) are selected that satisfy a trigger requiring two leptons, which may be electrons or muons. The invariant mass requirement, $m_{ll}$ $\ge$ $12$ GeV, has been applied for any opposite sign same-flavor (OSSF) pair of electrons and muons. This removes low-mass bound states and $\gamma^{*}$ $\rightarrow$ $l^{+}l^{-}$ production. It is required that at least one electron or muon in each event has a transverse momentum of $p_{T}$ $>$ $20$ GeV. Additional electrons and muons must have $p_{T}$ $>$ $10$ GeV and all of them must be within in the pseudorapidity of |$\eta$| $\le$ $2.4$. Tau leptons decay either into a lepton (electron or muon) and neutrinos or a hadronic final state generally made up of charged pions and neutral pions. The hadronic decays yield either a single charged track (one-prong) or three charged tracks (three prong) occasionally with additional electromagnetic energy from neutral pion decays. Both one- and three-prong candidates are used in this analysis if they have $p_{T}$ $>$ $20$ GeV. Leptonically decaying taus are included with other electrons and muons. Jets are reconstructed from all of the particle flow candidates using an anti-$k_{T}$ algorithm with a distance parameter of $0.5$, that have |$\eta$| $\le$ $2.5$ and $p_{T}$ $>$ $30$ GeV. Jets are required to have a distance $\Delta$R > 0.3 away from any isolated electron, muon, or $\tau_{h}$ candidate.

The background composition, arising from processes that produce genuine multilepton events, can be generally divided into two main sources. The most significant contributions to multilepton signatures are WZ and ZZ production, but rare processes such as $t\bar{t}$W and $t\bar{t}$Z can also contribute.  The second source are misidentified leptons, which can be classified in the following three categories: misidentified light leptons, misidentified $\tau_{h}$ leptons, and light leptons originated from asymmetric internal conversions, where a virtual photon decays promptly to a lepton pair and only one lepton passes the selection criteria. 

The contribution of misidentified light leptons can be estimated by measuring the number of isolated tracks and applying a scale factor between isolated leptons and isolated tracks. The $\tau_{h}$ misidentification rate is measured in a jet-dominated control sample by using a ratio the number of $\tau_{h}$ candidates in the signal region defined by $E_{cone}$ $<$ 2 GeV with respect to non-isolated $\tau_{h}$ candidates in  $6$ GeV $<$ $E_{cone}$ $<$ 30 GeV.

\begin{figure}[htbp]
\centering
\includegraphics[height=5cm]{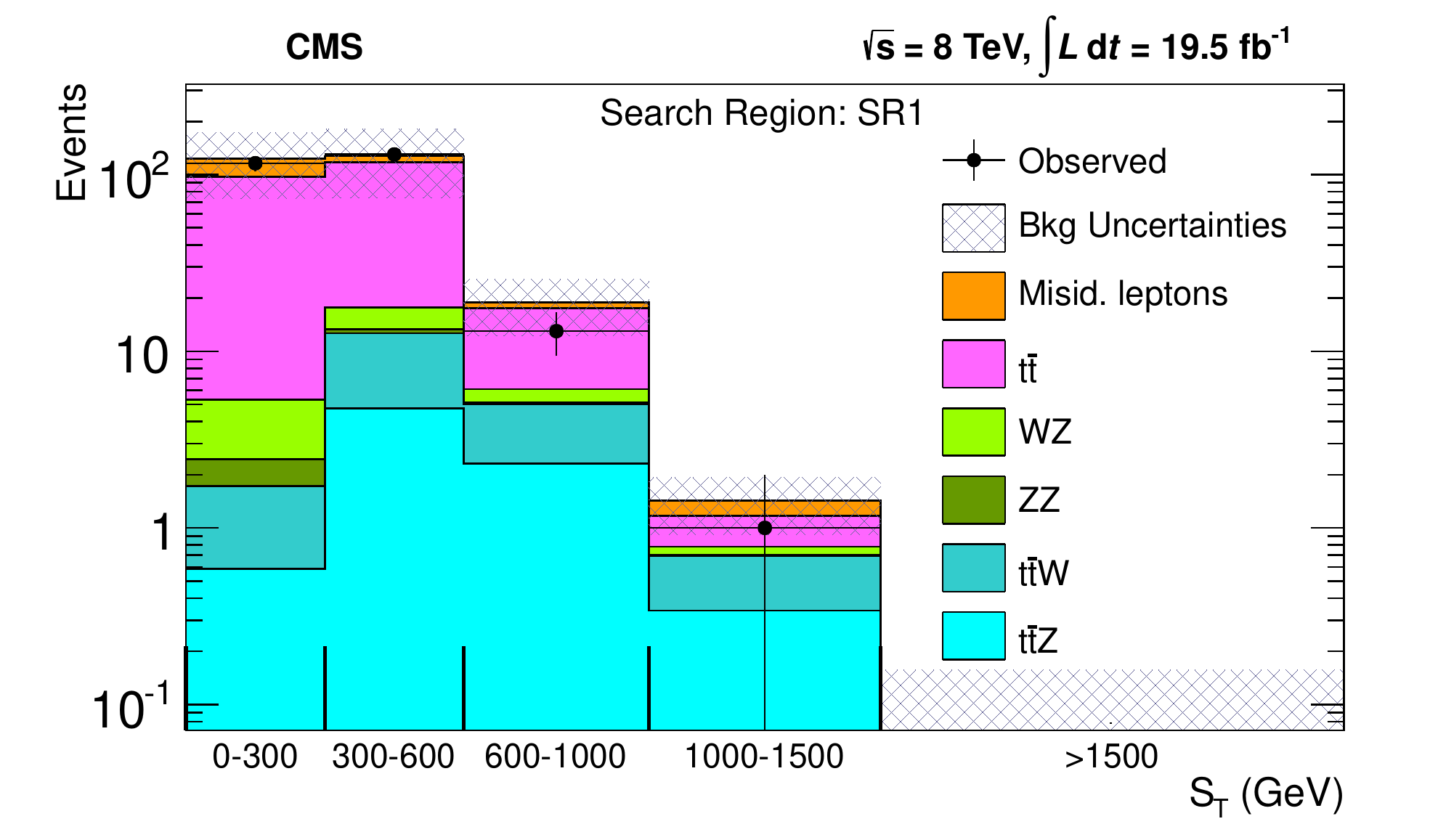}
\caption{The $S_T$ distribution for three lepton and b-quark jet events (SR1) including observed yields and background contributions. Both statistical and systematical uncertainties are shown in the shaded zone. The variable $S_T$ is the scalar sum of missing transverse momentum over all jets and isolated leptons.}
\label{fig:first}
\end{figure}

The rate of asymmetric conversion to light leptons is measured in a control region where no new physics expected. It is measured as the ratio of $l^{+}l^{-}l^{\pm}$ with respect to $l^{+}l^{-}l^{\gamma}$ candidates in the Z boson decays.

Depending on the total number of leptons, the number of $\tau_{h}$ candidates and whether there is a b tagged jet in the event. Eight signal regions are defined in five $S_T$ bins. The $S_T$ distribution for one of the signal region (SR$1$) is shown in Fig.~\ref{fig:first}. Data are in good agreement with the SM predictions in all signal regions.
\begin{figure}[t]
\centering
\includegraphics[height=10cm]{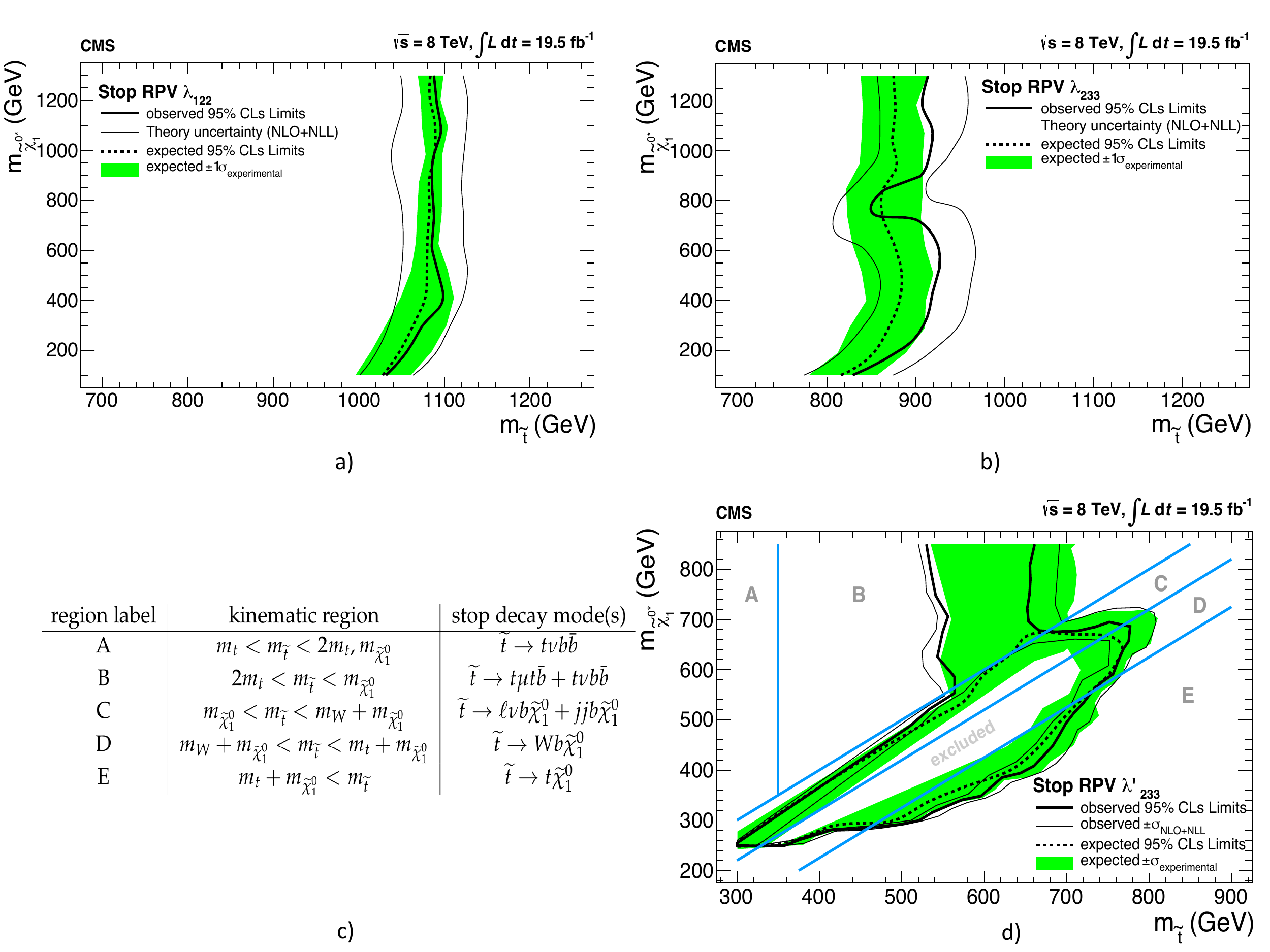}
\caption{The $95\%$ CL level limits in the stop mass and bino mass plane for models with RPV couplings $\lambda_{122}$(a), $\lambda_{233}$(b) and $\lambda^{`}_{233}$(d). For leptonic RPV couplings (a and b), the region to the left of the curve is excluded. For semileptonic RPV coupling (d), the region inside the curve is excluded. The kinematic properties of different regions for the $\lambda^{`}_{233}$ exclusion result from different stop decay products as explained in Table (c).}
\label{fig:second}
\end{figure}

To demonstrate the sensitivity for various signal-model scenarios for RPV couplings, the light decays to a top quark and intermediate on- or off-shell bino ($\tilde{t_1}$ $\rightarrow$ $\tilde{\chi_{1}}^{0*}+t$) is discussed in Fig.~\ref{fig:second}. The bino then decays to two leptons and a neutrino through the leptonic RPV interactions or through the semileptonic RPV interactions. The stop is assumed to be right-handed, and the RPV couplings are large enough that all decays are prompt. In the leptonic RPV SUSY, where $\lambda_{ijk}$ $\neq$ $0$, the corresponding limits are approximately independent of the bino mass and the stop mass below $1020$ GeV and $820$ GeV are excluded for $\lambda_{122}$ and $\lambda_{233}$, respectively. For the $\lambda_{233}$ coupling there is a change kinematics at the $m_{\tilde{\chi}^{0}_{1}}$ = $m_{\tilde{t}_{1}}$ - $m_{t}$, which below the stop decay is two-body, while above it is a four-body decay. In the region, around $\sim$$750$ GeV, the $\tilde{\chi}^{0}_{1}$ and top are produced at rest, which results in soft leptons, reducing the acceptance. For semileptonic coupling, which has non-zero  $\lambda^{'}_{233}$, the kinematics of the decay are more challenging. These different kinematic regions are shown in Fig.~\ref{fig:second}. The most significant effects, happens where $\tilde{\chi}^{0}_{1}$ $\rightarrow$ $\mu$+$t$+$b$ is kinematically disfavoured, as can be seen in region B, where the number of available leptons is reduced. The regions, where this effect is pronounced drive the shape of the exclusion for $\lambda^{'}_{233}$. The observed limit is stronger than the expected one so that it allows the observed exclusion region to reach into the regime where the bino decouples.

\section{Search for RPV SUSY in the four-lepton final state}

In this analysis, the lepton number violating term ($\lambda_{ijk} L_{i} L_{j}\bar{e}_k$), which causes the LSP in SUSY model to decay into four leptons, is studied~\cite{RPV2}. The main goal of this analysis is that the RPV term exists on top of some underlying RPC model, with properties which are currently barely constrained. Therefore, the results are interpreted by exploring RPV on top of very specific RPC SUSY pMSSM model in addition to the simplified model approach. 

Events are selected with at least one electron or muon with transverse momentum $p_T$$>$$17$ GeV, and another electron or muon with $p_T$$>$$8$ GeV which satisfies the trigger requirement. Events are reconstructed using the particle flow algorithm approach. It is required that leading highest electron or muon has $p_T$$>$$20$ GeV. Additional electrons or muons must have $p_T$$>$$10$ GeV and all of them must be within |$\eta$| $\le$ $2.4$. In order to remove quarkonia resonances, photon conversions, and low-mass continuum events the $m_{ll}$ $\ge$ $12$ GeV invariant mass cut, which is discussed in the previous section, is applied. Events with exactly 4 isolated leptons (electron and/or muons) containing at least one OSSF pair is selected. And then all OSSF lepton pairs with an invariant mass closest to the Z mass of $91$ GeV are determined. The invariant mass of this lepton pair and  the remaining lepton pair are defined in 2 dimensional distribution. Each mass are then classified as "below Z mass" (M $<$ $75$ GeV), "in Z mass" ( $75$ GeV $<$ M $<$ $105$ GeV) and "above Z mass" (M $>$ $105$ GeV). This provides nine regions reflecting different kind of resonant and non-resonant $4$-lepton production.  The presence of 4 prompt leptons, which is the only selection applied to the data in this analysis, is sufficiently discriminating on its own. The SM processes contributing to this signature are processes producing exactly 4 prompt or more leptons (ZZ, Z$t\bar{t}$, WW$t\bar{t}$, WWZ, WZZ and ZZZ), processes producing 3 prompt and one non-prompt lepton (WZ and W$t\bar{t}$) and Drell-Yan production with two extra non-prompt leptons. The contribution of non-prompt leptons is estimated using the fake rate technique, which is extensively explained in the public note. Consequently, the observed number of events in different background processes are consistent with the SM background expectations.   
\begin{figure}[b]
\centering
\includegraphics[height=8cm]{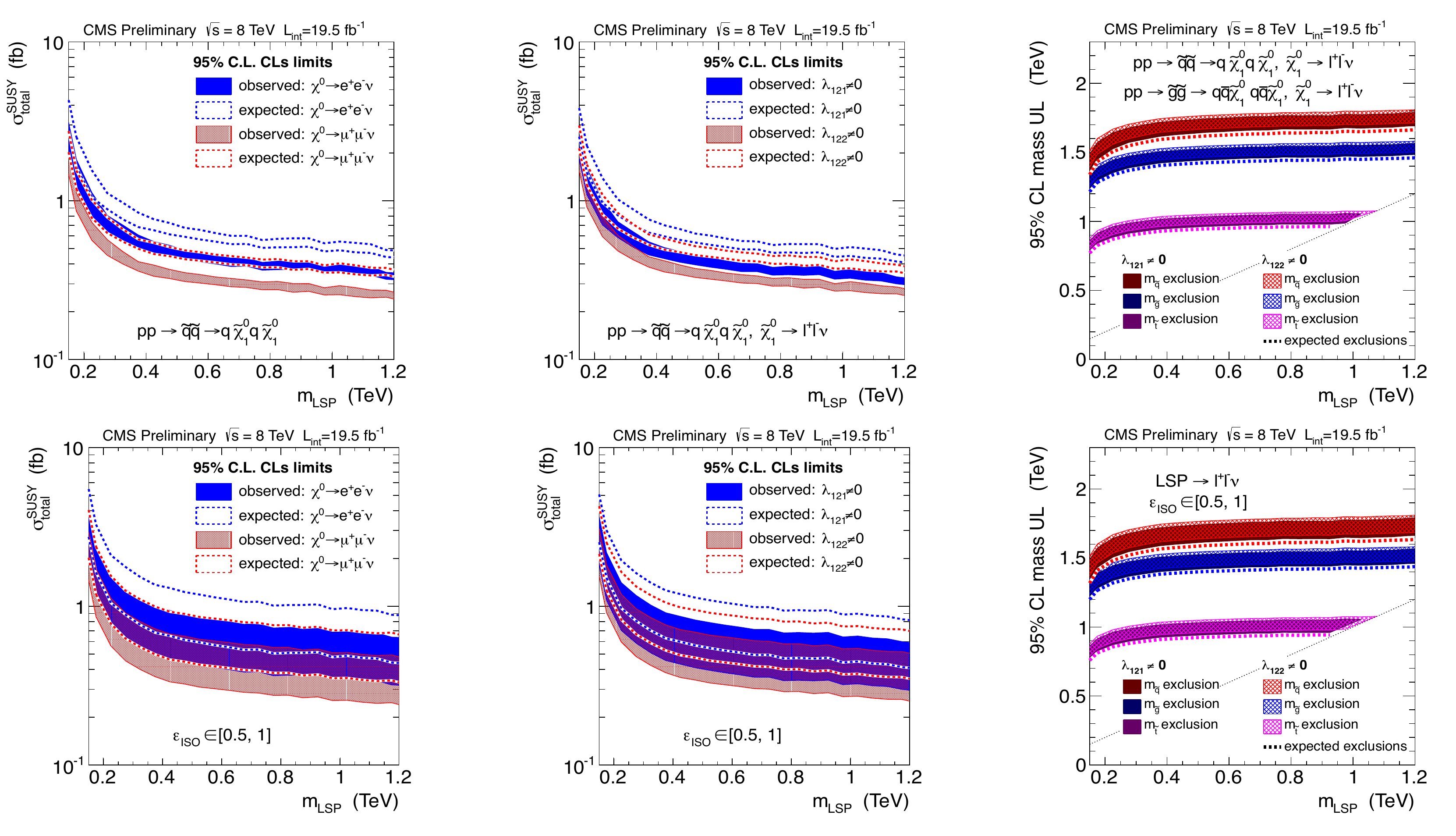}
\caption{$95$$\%$ C.L. upper limit on the mass and cross section of the simplified models (upper row) and generic SUSY models (lower row). Each band corresponds to the isolation efficiency for each SUSY models[cite]. The middle column shows the result for neutralino decaying exclusively to electrons or muons. The right column shows the result for the lepton flavors mixture corresponding to $\lambda_{121}$ and $\lambda_{122}$. A $30$$\%$ theoretical uncertainty for NLO+NLL calculations of SUSY production cross sections is included in the uncertainty band.}
\label{fig:third}
\end{figure}
One of the features of this analysis is the determination of the lepton efficiency for neutralino decays. The kinematics of these leptons are in general driven by the momentum distribution of the decaying neutralinos and their mass. In most scenarios, simplified models as well as generaic SUSY models, the lepton momentum is well above threshold, which results in high efficiency. However, large hadronic activity in the event can generally reduce the isolation efficiency. Therefore, it is concluded that the reduction of the total efficiency for this search may be up to $50$$\%$. As a result, once an upper limit $\sigma$x$L$x$\epsilon$ is extracted from the observations, and the efficiency is evaluated, the corresponding limit on the cross section, $\sigma_{total}^{SUSY}$, may be calculated. 

The cross section and mass exclusion limits are presented in Fig.~\ref{fig:third} for simplified and generic SUSY models. Using the total cross sections as a function of the mass of the corresponding SUSY particles, the cross section limit bands into mass exclusion bands as a function of the LSP mass is presented. Results for neutralinos decaying exclusively to electrons and muons and an appropriate mixture of electrons and muons in neutralino decays are also shown. In the analysis it is discussed that the kinematic efficiency is controlled by the neutralino mass and only weakly depends on the neutralino momentum. For the cases, where $\lambda_{121}$ or $\lambda_{122}$ has non-zero RPV coupling, the gluino mass is generally excluded below about $1.4$ TeV for a neutralino mass higher than $400$ GeV in case of $\lambda$ sufficiently large decay to prompt neutralino decays. For the benchmark point considered with a $2.4$ TeV gluino, squarks with a mass below about $1.6$ TeV are excluded.   
\section{Summary}
Results of searches for RPV SUSY in events with multilepton final states at the CMS experiment have been presented. The final number of events selected in data are consistent with the predictions for SM processes and no evidence of SUSY has been observed. The results of the leptonic RPV SUSY $\lambda_{ijk}$ and semileptonic RPV SUSY $\lambda^{'}_{ijk}$ searches are discussed in the context of the pMSSM and various simplified models. In the absence of signal, limits on the allowed parameter space in the corresponding models are set. In addition, a new approach for interpreting experimental observations are discussed in the pMSSM framework, allowing for a more general conclusion possible for SUSY searches.


\end{document}